# Spin gapless semiconducting behavior in equiatomic quaternary

# CoFeMnSi Heusler alloy

Lakhan Bainsla,[1] A. I. Mallick,[1] M. Manivel Raja,[2] A. K. Nigam,[3] B.S.D.Ch.S. Varaprasad,[4]

Y. K. Takahashi,[4] Aftab Alam,[1] K. G. Suresh[1,#] and K. Hono[4]

[1]Department of Physics, Indian Institute of Technology Bombay, Mumbai 400076, India
[2]Defence Metallurgical Research Laboratory, Hyderabad 500058, India
[3]Department of Condensed Matter and Materials Science, Tata Institute of Fundamental Research, Mumbai 4000005
[4]Magnetic Materials Unit, National Institute for Materials Science, Tsukuba 305-0047, Japan

## Abstract

Spin gapless semiconductors (SGS) form a new class of magnetic semiconductors, which has a band gap for one spin sub band and zero band gap for the other, and thus are useful for tunable spin transport based applications. In this paper, we report the first experimental evidence for spin gapless semiconducting behavior in CoFeMnSi Heusler alloy. Such a behavior is also confirmed by first principles band structure calculations. The most stable configuration obtained by the theoretical calculation is verified by experiment. The alloy is found to crystallize in the cubic Heusler structure (LiMgPdSn type) with some amount of disorder and has a saturation magnetization of 3.7 $\mu_B$/f.u. and Curie temperature of 620 K. The saturation magnetization is found to follow the Slater-Pauling behavior, one of the prerequisites for SGS. Nearly temperature-independent carrier concentration and electrical conductivity is observed from 5 to 300 K. An anomalous Hall coefficient of 162 S/cm is obtained at 5 K. Point contact Andreev reflection data has yielded the current spin polarization value of 0.64, which is found to be robust against the structural disorder. All these properties are quite promising for the spintronic applications such as spin injection and can bridge a gap between the contrasting behavior of half-metallic ferromagnets and semiconductors.

Key words: spin gapless semiconductor, Heusler alloys, spin polarization, spintronics

PACS : 85.75.-d, 75.47.Np, 72.20.-i, 76.80.+y, 75.76.+j

Corresponding author (#: suresh@phy.iitb.ac.in

In recent studies, a new class of materials known as spin gapless semiconductors (SGS) has been reported to be promising for spintronic applications [1-3]. SGS have a band structure in which one spin polarized sub-band resembles that of a semiconductor, while the other sub band has a zero band gap at the Fermi level (see Fig. 1 for schematics of Density of States). Therefore, they combine the band structures of a ferromagnet and a semiconductor. Because of their unique properties, these are being considered as substitutes for diluted magnetic semiconductors (DMS) and are receiving intense research interest since the theoretical prediction. The major drawback of DMS is their low Curie temperature [4,5], which can be overcome in many anticipated SGS materials like some Heusler alloys.

For spintronics, Heusler alloys have a special place due to their high Curie temperatures ($T_C$) and tunable electronic/magnetic properties [6]. Many Heusler alloys have been theoretically predicted to exhibit SGS behavior [3,7-9]. However, the inverse Heusler alloy, $Mn_2CoAl$ is the only material from the Heusler family in which the SGS behavior has been confirmed experimentally [2]. Very recently this alloy has been studied in the thin film form as well, in order to check the applicability of the material in devices [10-11].

In the case of half-metallic materials, the electronic structure is metallic for one channel and semiconducting for the other channel, and hence the electrical transport is governed by the electrons with only one type of spin. SGS possess an open band gap for one channel and closed gap for the other channel. When the top of the valence band and the bottom of the conduction band for majority electrons touch the Fermi level (Fig. 1(d)), the resulting structure gives rise to SGS properties. As it is close to the critical point of zero-gap, SGS band structure is very sensitive to external influences, e.g., pressure or magnetic field. The density of states (DOS) schemes for a typical normal metal, semiconductor, half-metallic ferromagnet (HFM) and a SGS are compared in the Fig. 1. Some of the unique properties in the case of SGS are: (i) spin polarized current resulting from electrons as well as spin polarized holes; (ii) ability to switch between *n* or *p* type spin polarized carriers by applying an electric field; (iii) almost no energy required to excite electrons from the valence band to the conduction band.

Conventional full (ternary) Heusler alloys, with the stoichiometric composition $X_2YZ$, where X and Y are the transition metals and Z is a sp (or main group) element, have the cubic $L2_1$ structure (space group Fm-3m). However, when two ternary Heusler alloys, $X_2YZ$ and $X_2'YZ$, are combined together to form a quaternary compound XX'YZ with the stoichiometry 1:1:1:1, they show the LiMgPdSn prototype or the so called Y-type structure (space group F-43m) with somewhat different symmetry. These quaternary Heusler alloys have been explored only very little for their functional properties. [8,12,13]. The compound CoFeMnSi (henceforth referred to as CFMS) can be regarded as the combination of $Co_2MnSi$ and $Fe_2MnSi$. In the present work, we report experimental evidence of SGS behavior in bulk CFMS alloy, supported by *ab-initio* calculations. The alloy has been studied using structural, magnetization, spin polarization, magneto-transport and Hall effect measurements to probe the SGS behavior.

The polycrystalline sample of the CFMS alloy was prepared by arc melting of stoichiometric quantities of constituent elements (at least 99.9% purity) in argon atmosphere. Sample was flipped and melted several times to increase the homogeneity and the final weight loss was less than 1%. To further increase the homogeneity, the as-cast sample was annealed under vacuum for 14 days at 1073 K and then quenched in cold water. The crystal structure was investigated by x-ray diffraction pattern (XRD) collected at room temperature using Cu $K_\alpha$ radiation. $^{57}$Fe Mössbauer spectra were collected at room temperature using a constant acceleration spectrometer with 25 mCi $^{57}$Co(Rh) radioactive source. The spectra were analyzed using PCMOS-II least-squares fitting program. Rectangular piece of 10x5x2 mm$^3$ has been taken out from the annealed sample for the point contact Andreev reflection (PCAR) measurements. Current spin polarization measurements were done using the PCAR technique [14]. Sharp Nb tips prepared by electrochemical polishing were used to make point contacts with the sample. Spin polarization of the conduction electrons was obtained by fitting the normalized conductance $G(V)/G_n$ curves to the modified Blonder-Tinkham-Klapwijk (BTK) model [15]. A 'multiple parameter least squares fitting' was carried out to deduce current spin polarization (*P*) using dimensionless interfacial scattering parameter (*Z*), superconducting energy gap (*Δ*) and *P* as variables. The magnetic and transport measurements were performed in the temperature range of 5–300 K and in fields up to 50 kOe, using the PPMS (Quantum Design).

First principles electronic structure calculations were performed with the spin-polarized density functional theory (DFT) within Vienna *ab-initio* simulation package (VASP) [16] with a projected augmented-wave basis [17] using generalized gradient approximation (GGA) exchange correlation functional. A Monkhorst-Pack Brillouin-zone integration with 24 x 24 x 24 **k**-mesh was used for the calculation. We have used high precision with large plane wave cutoffs (340 eV), giving convergence within $10^{-1}$ meV/cell (10 kBar) for energy (stress tensor). Experimental lattice constant (a = 5.658 Å) obtained in our work has been used for all the calculations.

In order to check the phase stability of the structure, we have calculated the site preference energies of various configurations for CFMS. The F-43m crystal structure (with atomic label ABCD) is shown in the inset of Fig. 2. The structure can be considered as the four inter-penetrating fcc sublattices with Wyckoff positions A(0,0,0), B(1/4,1/4,1/4), C(1/2,1/2,1/2) and D(3/4,3/4,3/4). Considering the symmetry of the structure, we fixed the position of Si-atom at D-sites and all combinations of the other three elements (in total 6 configurations) were checked. Out of 6 possible configurations, only four configurations are distinct and the rest two are energetically degenerate. The site preference energy of the four distinct configurations is shown in Fig. 2. $E_0$ is just a reference energy, which corresponds to the most stable configuration (Type-4). The two degenerate configurations are – Type-5: Co - A, Fe - B, Mn - C (equivalent to Type-3) and Type-6: Co - A, Fe - C, Mn - B (equivalent to Type-4). These calculations give an accurate idea about the most favorable occupancy scheme of the constituent atoms in CFMS. The calculated magnetic moments for the energetically most favorable configuration (i.e. Type-4) are: $\mu_{Co}$ = 0.82 $\mu_B$/atom, $\mu_{Fe}$ = 0.53 $\mu_B$/atom, $\mu_{Mn}$ = 2.72 $\mu_B$/atom. The total magnetic moment per cell is found to be $\mu_{Tot.}$ = 4.01 $\mu_B$, which follows the Slater-Pauling behavior with a simple linear scaling: $\mu_{Tot}$ = $Z_t$ – 24 where $Z_t$ is the total number of valence electrons. Experimentally, our structural analysis also revealed that the alloy exists in the Type-6 configuration (which is equivalent to Type-4).

Figure 3 shows the calculated spin polarized band structure and the density of states of the energetically most favorable configuration (Type-4 or Type-6). One can notice that the DOS exhibits a band gap ~0.62 eV (semiconducting behavior) in one spin sub band, while the Fermi level falls within a negligible energy gap in the other spin sub band. The minority spin band structure indicates a band gap between the $t_{2g}$ and $e_g$ states, which resembles the band structure of other nonmagnetic semiconductors. Notably, the band structure for the majority spin band shows an indirect nature of the gap, where the valence band at Γ almost touches the conduction band at X, which corresponds to a deep valley in the DOS at $E_F$. Such a nearly closed band gap character in one spin channel suggests CFMS to be a close spin gapless semiconductor.

The superlattice reflections were clearly observed in the XRD pattern, which reveal that the alloy exists in ordered cubic Heusler structure (LiMgPdSn type) with F-43m space group (no. # 216). Lattice parameter of the alloy was found to be a = 5.658 Å from the Rietveld refinement, which is in close agreement with the earlier report [12]. Further information about the crystal structure was derived with the help of $^{57}$Fe Mössbauer spectroscopic measurements at room temperature as shown in the Fig 4. The experimental Mössbauer spectrum has been fitted with three sextets and a doublet having hyperfine field ($H_{hf}$) values of 290, 132 and 98 kOe and relative intensities of 38, 35, 17 and 10% respectively. The quadrupole shift value is almost zero, which is in accordance with the cubic symmetry of local Fe environment. Three sextets $S_1$, $S_2$, $S_3$ and a doublet were found to be essential to obtain a good fit. The best fit of the spectrum was obtained by considering the type-6 configuration (which is equivalent to type-4) in LiMgPdSn structure, where Co, Mn, Fe and Si occupy A, B, C and D site respectively. For a well ordered LiMgPdSn structure, Fe atoms must occupy the Y sites with cubic symmetry ($O_h$) resulting in a single sextet because there is only one crystallographic site for Fe. The presence of three sextets indicates some amount of structural disorder in the alloy. A large decrease in $H_{hf}$ is expected when Fe occupies X and X' sites because these sites have the highest number of non-magnetic near neighbors; $H_{hf}$ is expected to not change much when Fe occupies the Z site as the number of magnetic near neighbors are similar for Y and Z sites. The experimentally observed value of $H_{hf}$ clearly indicates that Fe also occupies X, X' sites, resulting in DO$_3$ type disorder. The sub-spectra $S_1$, $S_2$ and $S_3$ are ascribed to the ordered LiMgPdSn phase, DO$_3$ (Y-site) and DO$_3$ (X-site) phases respectively. The intensity of $S_1$ (38%) is found to be higher as compared to $S_2$ (35%) and $S_3$ (17%), which implies that the structure is reasonably ordered at room temperature. The structural stability of the alloy was checked using differential thermal analysis (DTA) in the temperature range of 400 K to 1450 K. The DTA plot shows no structural transitions, indicating the structural stability in the entire temperature range investigated. A minimum in the DTA exothermic curves was observed near the Curie temperature ($T_C \approx 620$ K), which is in agreement with earlier observation of 623 K [12].

The temperature dependencies of the electrical conductivity [$\sigma_{xx}(T)$] and the carrier concentration [$n(T)$] are shown in Fig.5. $\sigma_{xx}(T)$ is measured under 0 and 50 kOe fields in the temperature range of 5- 300 K. Electrical conductivity increases with increase in temperature, indicating a non-metallic conduction. $\sigma_{xx}(T) = 2980$ S/cm and $\sigma_{xx}(T) = 3000$ S/cm are obtained at temperature of 300 K under the field of 0 kOe and 50 kOe, respectively. The electrical conductivity value obtained at 300 K for CFMS is slightly higher than that reported for Mn$_2$CoAl (i.e. 2440 S/cm) [2]. $\sigma_{xx}(T)$ varies linearly in the high temperature region, while a non-linear behavior is observed in the low temperature region. The disorder-enhanced coherent scattering of conduction electrons may be the cause for the non-linear behavior at low temperatures [18]. The conductivity behavior is unusual and different from that of the normal metals or semiconductors. Carrier concentration of the CFMS alloy is calculated from the Hall coefficient ($R_H$) measurements. The temperature independent carrier concentration ($n$) is observed in the temperature range from 5 to 300 K, which is typical of spin gapless semiconductors [2,19]; carrier concentration of 4 x $10^{19}$ cm$^{-3}$ is observed at 300 K, which is in between the value observed for HgCdTe ($10^{15}$ – $10^{17}$ cm$^{-3}$), Mn$_2$CoAl ($10^{17}$ cm$^{-3}$) and Fe$_2$VAl ($10^{21}$ cm$^{-3}$) [20]. The behavior of $\sigma_{xx}$(T) and $n$(T) strongly supports the SGS behavior in the material. The observations of low carrier concentration and high resistivity show the exceptional stability of electronic structure and its insensitivity to the structural disorder in this alloy.

The anomalous Hall conductivity $\sigma_{xy} = \rho_{xy}/\rho_{xx}^2$ at 5 K was obtained from the magnetic field dependent transport measurements (as shown in Fig. 6) in order to study the low field behavior in more detail. Hall conductivity follows the same behavior as observed for the magnetization isotherm (inset) as shown in the Fig. 6. The anomalous Hall conductivity ($\sigma_{xy0}$) value is defined as the difference in $\sigma_{xy}$ values at zero and the saturation fields. It is found that $\sigma_{xy0}$ attains a value of 162 S/cm, which is higher than that observed in Mn$_2$CoAl (22 S/cm) [2], but less than that of the half metallic Co$_2$FeSi ($\approx$ 200 S/cm at 300 K) [21], and Co$_2$MnAl ($\approx$ 2000 S/cm) [22]. As can be seen from the inset, the saturation magnetization (M$_S$) value of 3.7 $\mu_B$/f.u. is obtained at 3 K, which is less than the calculated value of 4.01 $\mu_B$/f.u. The structural disorder observed from the Mössbauer data may be the reason behind the disagreement between the observed and calculated values of M$_S$.

The current spin polarization (P) measurements at the ferromagnetic (FM)/superconductor (SC) point contact were done by using PCAR technique. All the conductance curves (Fig. 7) were recorded at the temperature of 4.2 K by using Nb as the superconducting tip. The normalized conductance curves were fitted to the BTK model [17] by keeping P, $\Delta$ and $Z$ as variables. Due to the absence of proximity effect in the PCAR data, we assumed $\Delta = \Delta_1 = \Delta_2$. The values of fitted parameters with best fitting (least $\chi^2$ values) are shown in the figures. The shape of the conductance curves depends on the value of Z, with the curves becoming flat near the $\Delta$ for low Z values. The values of $\Delta$ obtained from the best fit are lower than the bulk superconducting band gap of Nb = 1.5 meV, which is due to the multiple contacts at the interface [23]. The intrinsic value of current spin polarization is obtained by recording the conductance curve for Z = 0. However, in the present case, the lowest possible Z value was 0.10 and therefore, the P vs. Z plot was extrapolated to Z = 0. The current spin polarization value of 0.64 is deduced from the P vs. Z plot, which is comparable to the value obtained in some high spin polarization materials by using PCAR technique such as $Co_2Fe(Ga_{0.5}Ge_{0.5})$ with P= 0.69 ± 0.02 [24], $Co_2Fe(Al_{0.5}Si_{0.5})$ with P= 0.60 ± 0.01 [25]. We attribute this polarization to the few states available for the majority sub band at the Fermi level.

In conclusion, the equiatomic quaternary Heusler alloy of CFMS was investigated theoretically as well as experimentally. *Ab-initio* calculations predicted the SGS type of electronic structure with open gap for one spin sub band and closed (negligibly small) for the other spin band. The phase stability of the alloy was checked by calculating the site preference energies; Type-4 configuration (equivalent to Type-6) was found to be most stable state. Total magnetization from *ab-initio* calculation is found to follow the Salter-Pauling behavior, which is one of the prerequisites for SGS. In order to check the proposed SGS behavior, the alloy was investigated by means of structural, magnetic, magneto-transport, spin polarization and Hall measurements. The structural analysis reveals that the alloy exists in cubic Heusler structure (prototype LiMgPdSn) with Type-4 configuration, which is also confirmed from the *ab-initio* calculation. $M_S$ value of 3.7 $\mu_B$/f.u. is obtained at 3 K, with Curie temperature of≈ 620 K. Nearly temperature -independent carrier concentration and the low values (~ $10^{-19}$ $cm^{-3}$) along with the electrical conductivity of the order of 3x$10^{-3}$ S/cm reveal that CFMS shows SGS behavior. The current spin polarization value of 0.64 is deduced from the PCAR conductance curves at 4.2 K. The above mentioned magnetic and transport properties of this novel SGS material seem to be very promising for the spintronic applications such as spin injection into semiconductors. One of our immediate endeavors is to theoretically investigate the effect of antisite defects, which shows some evidence in our Mössbauer spectra.

**Acknowledgements**

One of the authors, LB, would like to thank UGC, Government of India for granting senior research fellowship (SRF). KGS thanks ISRO, Govt. of India for the financial assistance in carrying out this work. AIM acknowledges the support from TAP fellowship under SEED Grant (project code 13IRCCSG020). The authors thank D. Buddhikot for his help in the resistivity and Hall measurements.

**Figure captions:**

FIG. 1. Schematic of density of states for a typical (a) metal, (b) semiconductor (c) half metal and (d) spin gapless semiconductor

FIG. 2. Site preference energies for different configurations of CoFeMnSi. Inset: crystal structure of prototype LiMgPdSn.

FIG. 3.  Band structure and density of states of CoFeMnSi: (a) majority-spin bands, (b) density of states, (c) minority-spin bands.

FIG. 4. [57]Fe Mössbauer spectrum of CFMS collected at room temperature.

FIG. 5. Temperature dependence of the electrical conductivity, $\sigma_{xx}(T)$ (left hand scale). Variation of carrier concentration, $n(T)$ with temperature (right hand scale).

FIG. 6. Field dependence of anomalous Hall effect (AHE) at 5 K in CFMS. The Hall conductivity, $\sigma_{xy}(T)$ is shown as the function of applied field. Inset shows the magnetization isotherm obtained at 5 K.

FIG. 7. Normalized conductance curves recorded at 4.2 K. In fig. (a) open circles denote the measured experimental data and solid lines are the fit to the data by using modified BTK model. Fig. (b) Represents the linear fit to P vs. Z data with extrapolation down to Z = 0.

**Figures:**

FIG. 1

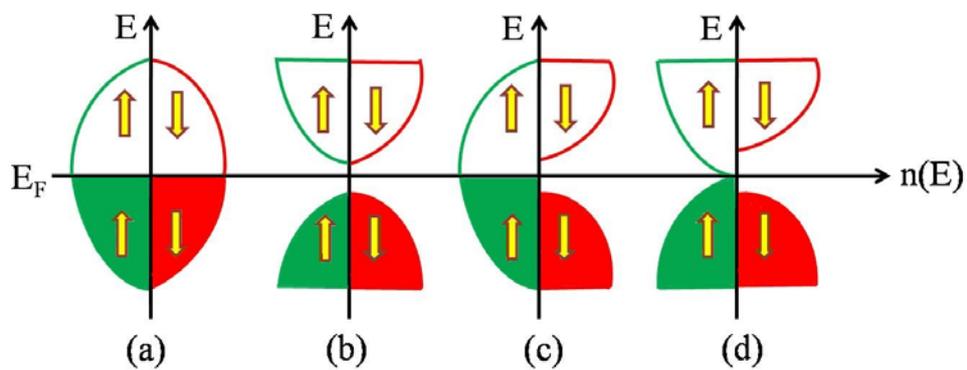

FIG. 2

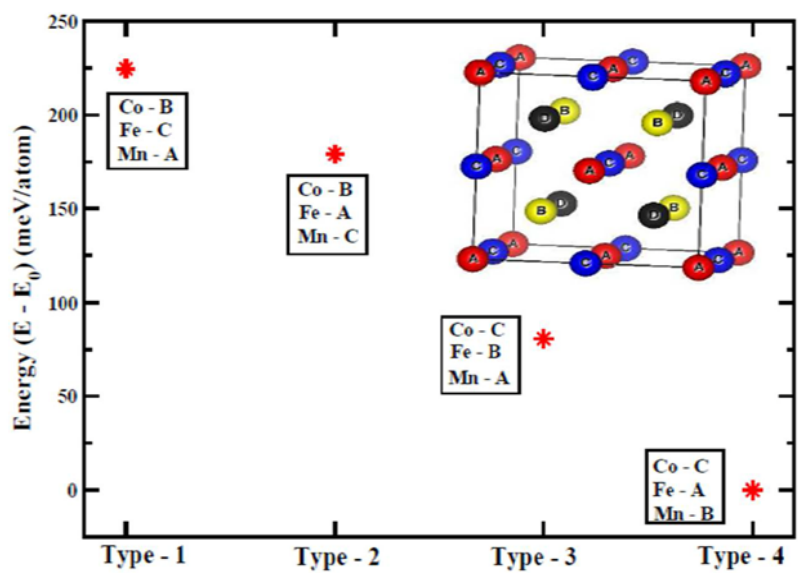

FIG. 3

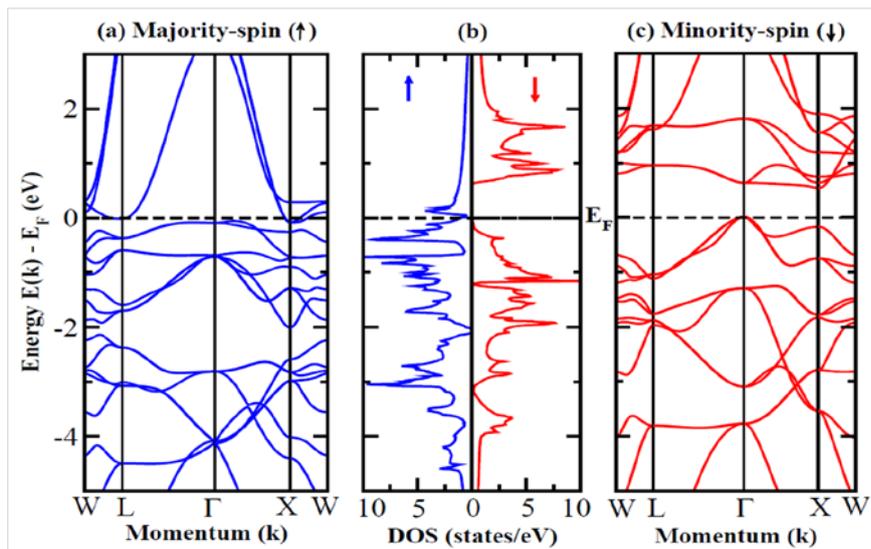

FIG. 4

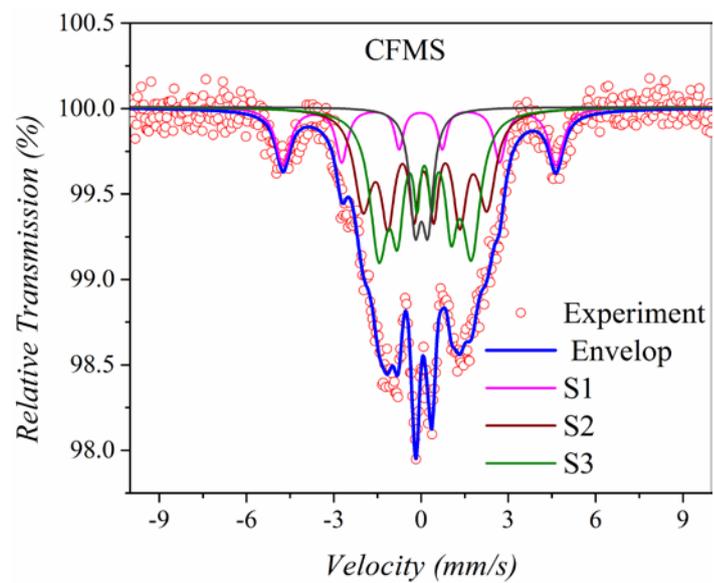

FIG. 5

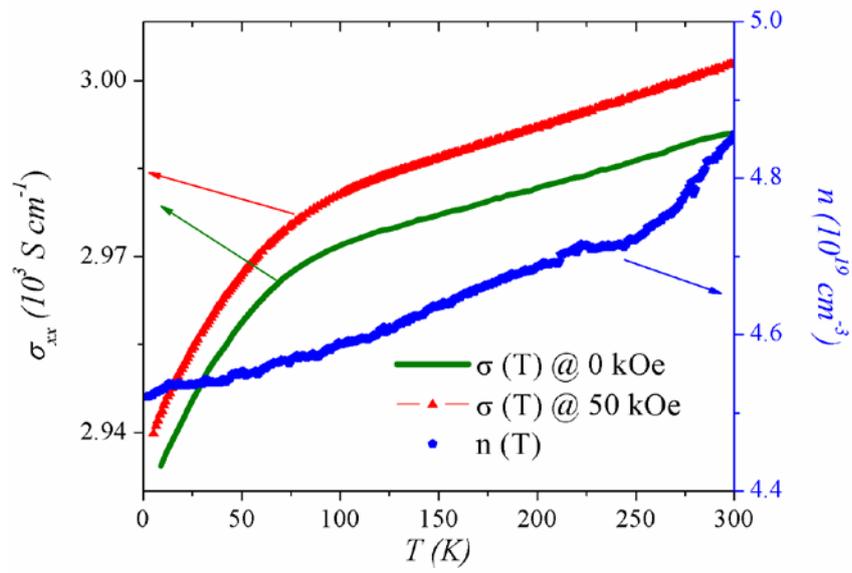

FIG. 6

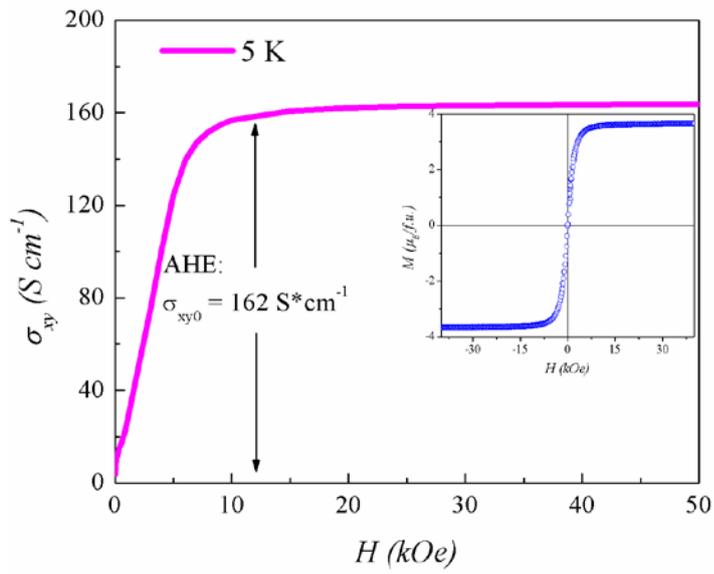

FIG. 7

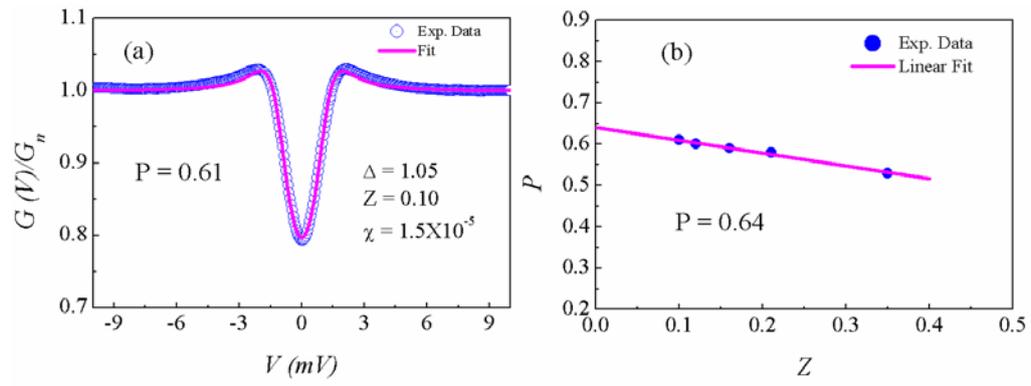